\documentclass[preprint,showpacs,superscriptaddress,preprintnumbers,longbibliography]{revtex4-1}
\usepackage{graphicx,bm,amssymb,amsmath,xcolor,pifont,soul,verbatim}
\usepackage{longtable,booktabs,multirow}
\begin{document}

\title{Materials design principles towards high hole mobility learning from an abnormally low hole mobility of silicon}

\author{Qiao-Lin Yang}
\affiliation{State Key Laboratory of Superlattices and Microstructures, Institute of Semiconductors, Chinese Academy of Sciences, Beijing 100083, China}
\affiliation{Center of Materials Science and Optoelectronics Engineering, University of Chinese Academy of Sciences, Beijing 100049, China}

\author{Hui-Xiong Deng}
\affiliation{State Key Laboratory of Superlattices and Microstructures, Institute of Semiconductors, Chinese Academy of Sciences, Beijing 100083, China}
\affiliation{Center of Materials Science and Optoelectronics Engineering, University of Chinese Academy of Sciences, Beijing 100049, China}

\author{Su-Huai Wei}
\affiliation{Beijing Computational Science Research Center, Beijing 100193, China}

\author{Shu-Shen Li}
\affiliation{State Key Laboratory of Superlattices and Microstructures, Institute of Semiconductors, Chinese Academy of Sciences, Beijing 100083, China}
\affiliation{Center of Materials Science and Optoelectronics Engineering, University of Chinese Academy of Sciences, Beijing 100049, China}

\author{Jun-Wei Luo}
\email{jwluo@semi.ac.cn}
\affiliation{State Key Laboratory of Superlattices and Microstructures, Institute of Semiconductors, Chinese Academy of Sciences, Beijing 100083, China}
\affiliation{Center of Materials Science and Optoelectronics Engineering, University of Chinese Academy of Sciences, Beijing 100049, China}
\affiliation{Beijing Academy of Quantum Information Sciences, Beijing 100193, China}

\newcommand{\mobun}{{cm$^2$/Vs}} 
\def\a{{\alpha}}
\def\b{{\beta}}
\def\ve{{\varepsilon}}
\def\w{\omega}
\def\bk{{\bf k}}
\def\bq{{\bf q}}
\def\bv{{\bf v}}
\def\bE{{\bf E}}
\def\bJ{{\bf J}}
\def\d{\delta}
\def\>{\rangle}
\def\<{\langle}
\def\D{\partial}
\def\kt{k_{\rm B}T}

\date{\today}

\begin{abstract}
{\bf Si dominates the semiconductor industry material but possesses an abnormally low room temperature hole mobility (505 \mobun), which is four times lower than that of Diamond and Ge ($\sim 2000$ \mobun), two adjacent neighbors in the group IV column in the Periodic Table. In the past half-century, extensive efforts have been made to overcome the challenges of Si technology caused by low mobility in Si. However, the fundamental understanding of the underlying mechanisms remains lacking. Here, we theoretically reproduce the experimental data  for conventional group IV and III-V semiconductors without involving adjustable parameters by curing the shortcoming of classical models. We uncover that the abnormally low hole mobility in Si originating from a combination of the strong interband scattering resulting from its weak spin-orbit coupling and the intensive participation of optical phonons in hole-phonon scattering. In contrast, the strong spin-orbit coupling in Ge leads to a negligible interband scattering; the strong bond and light atom mass in diamond give rise to high optical phonons frequency, preventing their participation in scattering.  Based on these understandings rooted into the fundamental atomic properties, we present design principles for semiconducting materials towards high hole mobility. }

\end{abstract}

\maketitle

Carrier mobility is one of the fundamental parameters of semiconductor materials since it governs the operation speed through the transit time across the device, the circuit operating frequency or the sensitivity in magnetic sensors~\cite{sze_physics_2007,chuang_physics_2009,luque_lopez_handbook_2011,mahan_best_1996,novoselov_two-dimensional_2005,ishiwata_extremely_2013}. Increasing carrier mobility is one of the critical strategies to make microchips faster and more energy-efficient. Specifically, increasing carrier mobility can enhance the transistor drive current, which can, in turn, improve device performance or maintain performance and reduce power consumption~\cite{pillarisetty_academic_2011}. Because the hole mobility ($\mu_h=505$ \mobun) is significantly lower than the electron mobility ($\mu_2=1450$ \mobun) in Si~\cite{pillarisetty_academic_2011,madelung_physics_1982}, the device performance of p-type transistor (PMOS) operates only at about one third the performance of n-type transistor (NMOS)~\cite{goley_germanium_2014}. It stimulates tremendous efforts in past decades in seeking materials with high hole mobility to replace Si to substantially improve PMOS device performance to maximize microchip performance gains [7]. Comparing to Diamond and Ge, its two adjacent neighbors in group IV column of the Periodic Table that have more superior hole transport properties ($\mu_h=1800$ \mobun\, and $\mu_h=1900$ \mobun\, respectively)~\cite{noauthor_materials_nodate}, the hole mobility in Si is unusually low with an abnormal trend~\cite{adachi_properties_2005}. Although Si is the most well-documented semiconductor due to its dominance in the semiconductor industry, this abnormal trend in hole mobility is a long-standing puzzle. The underlying mechanisms remain ambiguous.

The lack of the fundamental understanding may due to the complexity of hole-phonon scattering in semiconductor valence bands~\cite{kranzer_mobility_1974}, which contain essential complications, including the band degeneracy, the warping of the heavy-hole (hh) band, the p-type symmetry of the hole wave function, the influence of the light holes, and interband scattering between the light hole (lh) and hh bands~\cite{kranzer_mobility_1974,wiley_lattice_1970,conwell_lattice_1959,bir_galvanomagnetic_1962,wiley_polar_1971,costato_hole_1972}. These complications require considerable modifications of the scattering models by incorporating many adjustable parameters such as deformation potentials for acoustic and optical phonons~\cite{kranzer_mobility_1974,wiley_polar_1971}. It is arguable whether these adjustable parameters accurately contain all the effects of the complications mentioned above~\cite{kranzer_mobility_1974,lawaetz_low-field_1968}, and a fortuitous agreement between the experimental values and simple model calculations may occur. 
For instance, experimentally measured hole mobility data can be fitted well both with and without polar-mode scattering for semiconductors~\cite{wiley_lattice_1970,wiley_polar_1971}. In this sense, one usually ascribed the observed variation in hole mobility for group-IV and group-III-V materials to small variations in material parameters rather than different scattering mechanisms~\cite{wiley_polar_1971}. A recently developed first-principles approach rooted in the Boltzmann transport equation seems to avoid such uncertainties~\cite{giustino_electron-phonon_2017,ponce_towards_2018,ponce_hole_2019}. However, results show~\cite{ponce_towards_2018} that the calculated hole mobilities for Si can vary by about 60\% (from 502 to 820 \mobun) depending on the levels of band structure theory used, illustrating the importance of accurate band structure to the carrier mobility. Furthermore, although first-principles calculations may own the advantage of treating transport problems in their full complexity, it is also desirable to gain more physical insights by examining the individual mechanisms affecting the hole transport properties. 

In this work, we aim at elucidating the atomistic mechanisms underlying the abnormally low hole mobility in Si, rather than presenting an advanced method with improved accuracy. We calculate the phonon-limited hole mobilities of group IV and III-V semiconductors without adjustable parameters by employing the semi-empirical hole-phonon scattering theory~\cite{kranzer_mobility_1974,yu_fundamentals_2010} in conjunction with experimental valence band parameters~\cite{vurgaftman_band_2001,madelung_physics_1982}, and tight-binding approximations~\cite{harrison_elementary_1999} to directly compute the adjustable parameters. The success of reproducing experimental data enables us to disentangle the contributions from various scattering mechanisms since the used method offers the advantage of providing physical insight into the problem at hand. We illustrate that, in comparison with Ge, the spin-orbit coupling (SOC) is so weak that hh-lh band splitting is tiny, causing strong interband hole-phonon scattering and the abnormal low hole mobility in Si. On the other hand, Diamond is even lighter but has a much stronger bond than Si, so it has a very high optical phonon frequency, which prevents it from participating in the hole scattering. It renders Diamond having high hole mobility. These new level understandings of the atomic mechanisms governing the hole mobility enable us to propose design principles to identify semiconducting materials with high hole mobility.

The carrier mobility $\mu$ is related to the average scattering time $\tau$ and conductivity effective mass$m_c^*$: $\mu=e\tau/m_c^*$ ~\cite{yu_fundamentals_2010}. In pure materials, $\tau$ is limited mainly by scattering of carriers by various types of phonons (lattice vibrations)~\cite{kranzer_mobility_1974,yu_fundamentals_2010,seeger_semiconductor_2010}. Specifically, carriers can interact with phonons via deformation potentials since the band edge’s energy variation is related to the lattice strain caused by atomic displacement due to acoustic phonons~\cite{shockley_energy_1950} and nonpolar optical phonons~\cite{herring_transport_1956}. The former is called acoustic (AC) deformation potential scattering, and the latter is nonpolar optical (NPO) deformation potential scattering. In polar semiconductors, carriers can also couple with optical phonons via the induced longitudinal electrical field~\cite{frohlich_electrons_1954} due to the macroscopic polarization field generated by the out-of-phase vibration [longitudinal-optical (LO) phonon] in electric dipoles tends to strongly couple with carriers, termed polar optical phonon (PO) scattering. An analogy of PO scattering to longitudinal-acoustic (LA) phonons is the piezoelectric scattering~\cite{meijer_note_1953}. Here, we adopt semi-empirical scattering modes~\cite{kranzer_mobility_1974,willardson_transport_1975} developed in the past forty years to calculate the hole mobility $\mu_\lambda$ limited by these four hole-phonon scattering mechanisms, respectively. We then obtain the total hole mobility according to Matthiessen’s Rule $\mu_h^{-1}=\sum_{\lambda=1}^{4} \mu_\lambda^{-1}$~\cite{hamaguchi_basic_2017}. Note that there are many band structure parameters involved in the semi-empirical scattering modes. Here, we use the Modified Beckea-Johnson (MBJ) potential~\cite{becke_simple_2006,tran_accurate_2009} implemented in the Vienna ab-initio simulation package (VASP)~\cite{kresse_efficiency_1996} to compute spin-orbit splitting energy for semiconductors under investigations. According to the tight-bonding model~\cite{harrison_elementary_1999}, we get reasonable deformation potentials~\cite{wiley_valence-band_1970,blacha_deformation_1984,harrison_electronic_1989}, phonon energy, elastic constants~\cite{harrison_electronic_1989,baranowski_bond_1984} and dielectric constants~\cite{samara_temperature_1983,burstein_c1_1965} to avoid the utilization of adjustable parameters as usual. The conductivity and density-of-states effective masses are obtained based on Luttinger parameters $\gamma_1$, $\gamma_2$, $\gamma_3$ suggested commonly for III-V materials~\cite{vurgaftman_band_2001} and group IV~\cite{madelung_physics_1982}.We also consider the temperature-dependence of density-of-states conductivity effective mass~\cite{linares_mobility_1979,barber_effective_1967}. Calculations are detailed in Supplemental Material, and related parameters are listed in Table S1-6. 

\begin{figure}
	\centering
	\includegraphics[width=0.5\columnwidth]{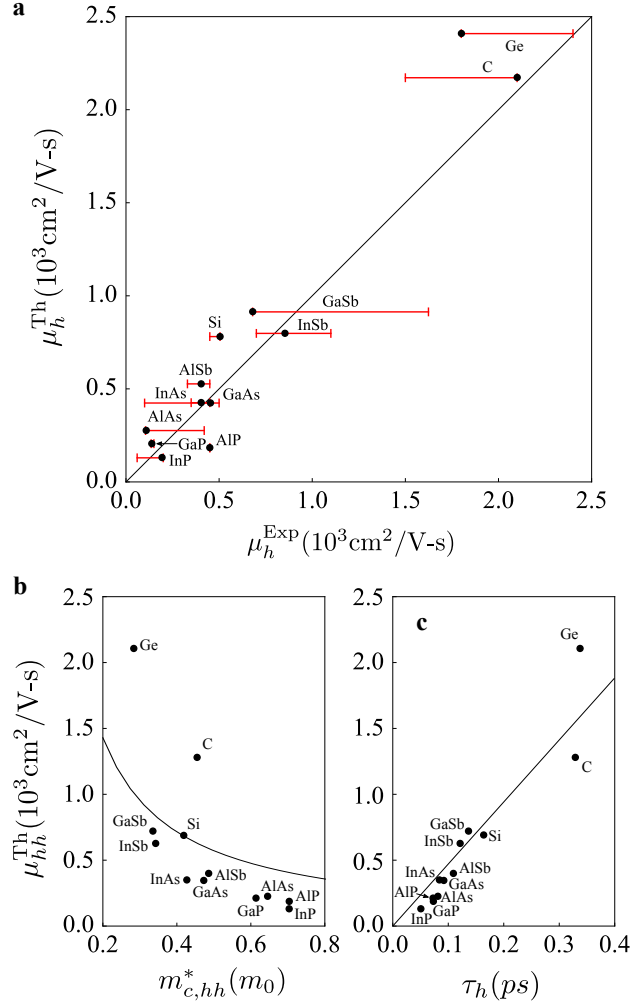}
	\caption{\label{fig1}
	{\bf The predicted hole mobility and hole-phonon scattering time for group III-V and group IV semiconductors at room temperature.}
		(a) Hole mobilities obtained from theoretical calculations $\mu_h^{Th}$ are plotted against the experimental values $\mu_h^{Exp}$ for group III-V and group IV semiconductors. Here, We use a red  line segment and a single dot, respectively, to indicate the range covered by multiple experimental values~\cite{madelung_physics_1982,noauthor_materials_nodate,adachi_properties_2005,noauthor_nsm_nodate} and the value cited in the Landolt-Bornstein data handbook on Semiconductors~\cite{madelung_physics_1982} for each material. The conductivity and density-of-states effective masses are obtained based on Luttinger parameters $\gamma_1$, $\gamma_2$, $\gamma_3$ suggested commonly for III-V materials~\cite{vurgaftman_band_2001} and group IV~\cite{madelung_physics_1982}. For Si, if we change the set of Luttinger parameters to that suggested by Ref.~\cite{ramos_structural_2001}, the predicted hole mobility becomes {580 \mobun} better agreement with experimental value. To avoid the uncertainty, we keep adopting the well-documented data in Refs. ~\cite{vurgaftman_band_2001} and ~\cite{madelung_physics_1982}.
		(b,c) The predicted heavy-hole mobility (b) against the conductivity effective mass of heavy-hole (hh) band, and (c) against the scattering time for group III-V and group IV semiconductors. The solid black lines are used to guide the eyes.
	}
\end{figure}

Fig.~\ref{fig1}(a) shows our calculated room-temperature hole mobilities against corresponding experimental data~\cite{madelung_physics_1982,noauthor_materials_nodate,adachi_properties_2005,noauthor_nsm_nodate} for group IV (Diamond, Si, and Ge) and group III-V (AlP, AlAs, AlSb, GaP, GaAs, GaSb, InP, InAs, and InSb) semiconductors. We use a range (marked by a red thick line) and single dot, respectively, to indicate multiple experimental values~\cite{madelung_physics_1982,noauthor_materials_nodate,adachi_properties_2005,noauthor_nsm_nodate} and the value cited in the standard semiconductor handbook~\cite{madelung_physics_1982} for each material. We find that the calculated values fall within the experimental range with a small scatter even though we adopted the semi-empirical hole-phonon scattering models. Specifically, the predicted Si hole mobility is {780 \mobun} against the suggested value of 505 \mobun~\cite{madelung_physics_1982}. If we change the set of Luttinger parameters to that suggested by Ref.~\cite{ramos_structural_2001}, the predicted value becomes {580 \mobun} in a be better agreement with experimental value. However, the first-principles predicted data of Si hole mobility ranges widely from 502 to 820 \mobun, depending on the functionals used in electronic band structure calculations~\cite{ponce_towards_2018}, Thus, our prediction exhibits reasonable accuracy. The reliability of the simple isotropic relaxation time model used here has already been confirmed by Fischetti and Laux~\cite{fischetti_band_1996}. They showed that the error introduced by neglecting the valence bands’ warping is less than 3\% for Si (which has high anisotropy) and is negligible for Ge. 

The extraordinary electron mobility of narrow band-gap semiconductors comes by virtue of their low effective mass $\mu_e^*$ ~\cite{ishiwata_extremely_2013}with the smallest band-gap InSb (0.17 eV at 300 K) possessing the highest ambient-temperature electron mobility (78,000 \mobun) of all known conventional semiconductors~\cite{durkop_extraordinary_2004}. However, the corresponding hole mobility in III-V semiconductors is orders of magnitude lower due to their considerably larger valence-band effective mass, together with their much shorter scattering time for hole transport than for electron transport~\cite{pillarisetty_academic_2011}. To assess the relative significance of effective mass and scattering rate in limiting the hole mobility, we have to separate the contributions of holes from hh, lh, and split-off (so) bands. Fig.S2 and Fig.S3 (Supplemental Material) shows that the hole mobility is enhanced only slightly by the admixture of lh holes due to its much higher mobility. In the rest of this work, we will focus on holes in the hh band.

Fig.~\ref{fig1}(b) shows the mobility against the hh conductivity mass for all considered semiconductors. We see that Ge and Diamond stand out clearly from remaining materials including Si. Specifically, Diamond has a similar effective mass as Si but is 2.5 times higher in (hh) hole mobility, demonstrating the hole-phonon scattering's primer role in hole mobility. All semiconductors except Ge and Diamond follow a clear overall trend that a lower hole effective mass possesses higher hole mobility, consistent with the common perception. Such an overall trend is also evidenced by the model mobility (solid line in Fig.~\ref{fig1}(b)) with an assumption of a constant hole scattering time (same as Si) but varying hole effective mass alone for all materials. It is expected for Ge and Diamond to have the superior hole mobility regarding non-polar group IV materials free from the polar scattering via optical phonons, which is a leading factor limiting the electron mobilities of group III-V semiconductors~\cite{kranzer_mobility_1974,yu_fundamentals_2010,boer_semiconductor_2018}. However, it is difficult to understand why non-polar Si shares the same trend as the polar group III-V semiconductors. 

An unrecognized, much stronger hole-phonon scattering causes this difficulty in Si relative to Diamond and Ge. We inspect the hole mobility against the scattering time for all considered semiconductors in Fig.~\ref{fig1}(c). It exhibits an excellent linear relationship between hole mobility and relaxation time for all considered semiconductors irrespective of their remarkable difference in hole effective masses. Such a linear relationship implies the hole-phonon scattering taking the role over the effective mass in determining the hole mobility. Fig.~\ref{fig1}(c) shows that Diamond and Ge have much longer scattering times $\tau$ than Si and all III-V materials. The short hole scattering time in group III-V semiconductors may arise from the polar scattering. However, as a non-polar semiconductor, Si possesses also a short hole scattering time, which is almost equal to that of polar semiconductors GaSb and InSb. Such short hole scattering time leads the hole mobility of Si to be only one-fourth of the hole mobilities of Ge and Diamond. The hole mobility of polar GaSb and InSb is even higher than that of non-polar Si due to their lighter hole effective masses. These exceptions imply complex hole-phonon scattering mechanisms in limiting the hole mobility in semiconductors. It is convenient to access the influence of each hole-phonon scattering mechanism ($\lambda$ = AC, NPO, PO, Pizo) on hole mobility since we deal with four scattering mechanisms separately. In non-polar group IV materials, the hole-phonon interactions are entirely governed by the deformation-potential scatterings due to both acoustic and optical phonons (particularly by longitudinal acoustic phonons (LA)). At room temperature, the valence band states occupied by holes are usually within $k_{B}T=25$ meV of the valence band maximum (VBM), which is at the $\Gamma$ point, so their wavenumbers tend to be pretty small. In this sense, the most critical phonons are those of small wavenumber or long wavelengths. For each scattering mechanism $\lambda$, we can also separate interband and intraband scattering rates. Subsequently, we could decompose the total scattering time $\tau$ into various scattering mechanisms $\tau_\lambda$ according to Matthiessen’s rule~\cite{hamaguchi_basic_2017},
  \begin{equation}\label{eq.1}
  \mu_h^{-1}=\sum_{\lambda} \mu_\lambda^{-1} 
  \end{equation}
\begin{figure}
	\centering
	\includegraphics[width=0.8\columnwidth]{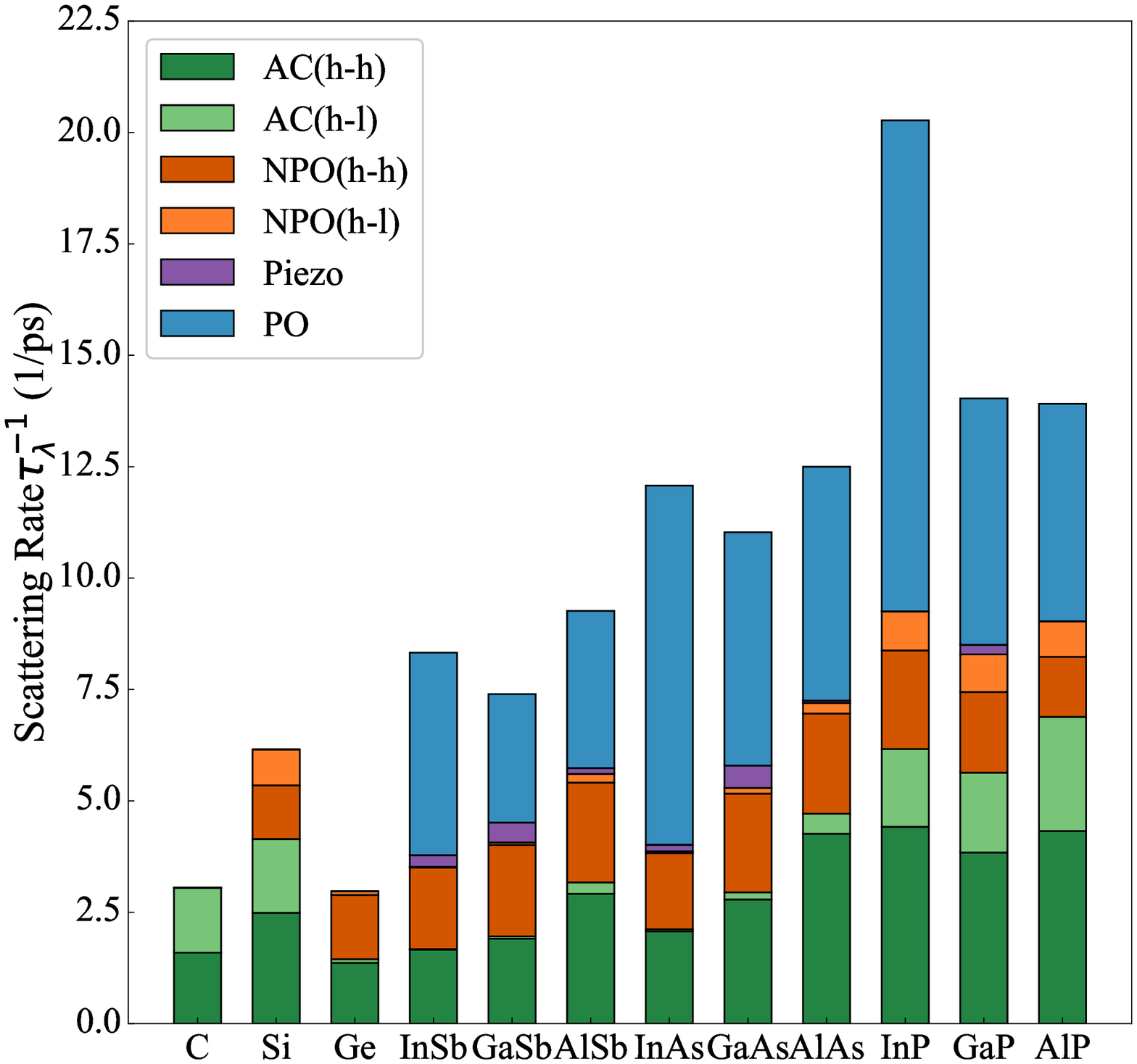}
	\caption{\label{fig2}
		 {\bf The scattering rate of the heavy hole for group IV and III-V semiconductors.} AC and NPO denote hole-phonon scattering via deformation potential by acoustic (AC) phonons and non-polar optical (NPO) phonons. The hh-hh and lh-hh in parentheses present intraband scattering within the hh band and interband scattering from hh band to lh band, respectively. Piezo and PO denote hole-phonon scattering via piezoelectric potential and polar electric potential induced by optical phonons, respectively. The piezoelectric potential scattering and polar optical scattering are called polar scattering. The interband transition is negligible for polar scattering in terms of Wiley’s theory~\cite{wiley_polar_1971}.
	}
\end{figure}

Fig.~\ref{fig2} displays the hole-phonon scattering rate $\tau_\lambda^{-1}$ of each scattering channel for all considered semiconductors. As expected, the piezoelectric scattering is absent in group IV semiconductors and is negligibly small relative to other scattering mechanisms~\cite{yu_fundamentals_2010} in all group III-V semiconductors. The sizeable polar scattering indeed yields polar group III-V semiconductors with much shorter total scattering time than non-polar group IV semiconductors, except Si. Among group III-V semiconductors, a significant variation occurs in the polar scattering rates: it plays a primary role in InAs, InSb, and InP (67\%, 55\%, and 54\%, respectively), but a minor role in GaSb and AlSb (less than 38\%). We find that the interband scattering channel substantially enhances the hole-phonon scattering in Diamond, Si, InP, GaP, and AlP. In Ge, however, such interband scattering is suppressed, rendering Ge has an almost three times longer scattering time $\tau$ than Si. From the comparison between Si and GaSb and InSb, we also see that the suppression of the interband scattering compensates the finite PO scattering in GaSb and InSb, yielding the total scattering times in GaSb and InSb almost the same as in Si. The reason makes Diamond having a three times longer scattering time than Si is, however, completely different since the Diamond benefits from the vanishing of NPO scattering and reduction in AC scattering rate. The different reasons resulting in the super hole mobilities in Diamond and Ge imply compelling physics limiting the hole mobility in semiconductors.

\begin{figure}
	\centering
	\includegraphics[width=0.8\columnwidth]{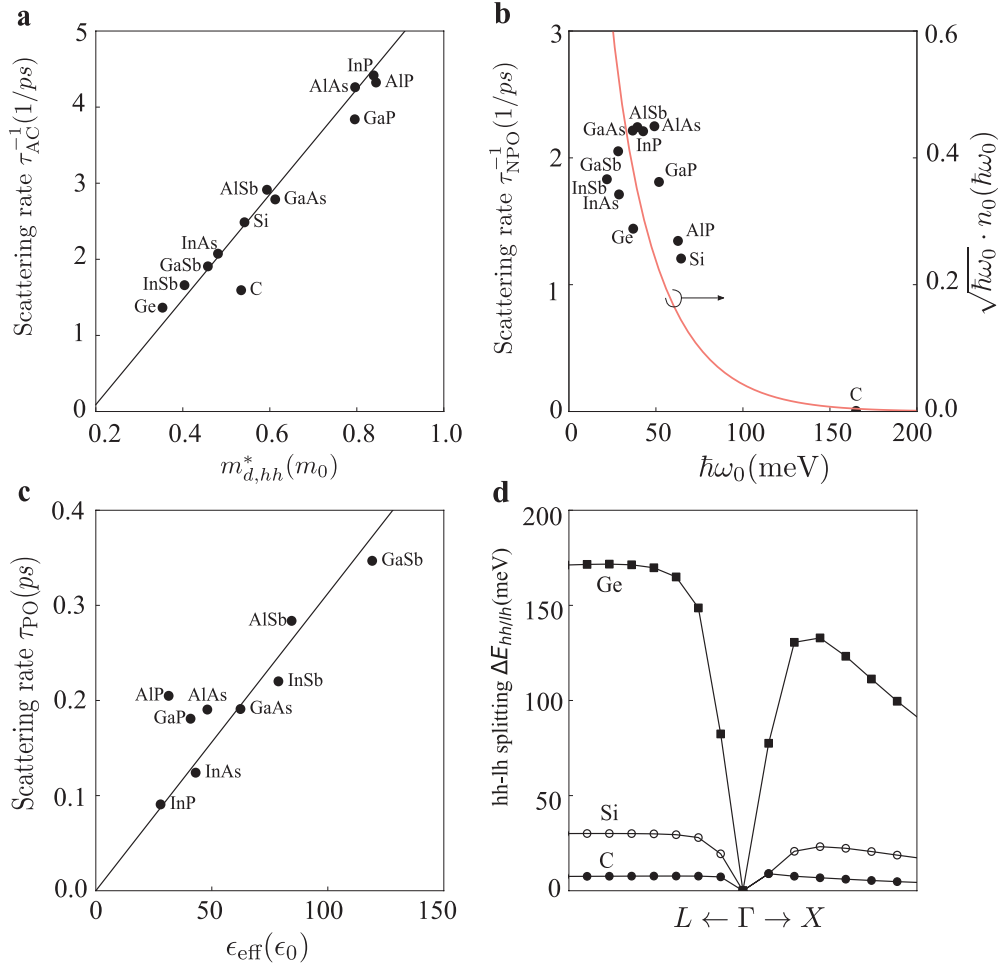}
	\caption{\label{fig3}
	{\bf Decomposition of hole-phonon scattering into a various factors, including DOS effective mass, optical phonon frequency, effective dielectric constant, and hh-lh splitting.}
		(a) the intraband AC scattering rate of group IV and group III-V semiconductors as a function of their DOS effective mass $m_{d,hh}$; 
		(b) the intraband NPO scattering rate and phonon occupation rate as a function of phonon energy, respectively; 
		(c) the PO scattering relaxation time as a function of effective dielectric constant $\varepsilon_{\textrm{eff}}$; 
		(d) The splitting energy $\Delta_{lh}$ between heavy- and light-hole band about Si, Ge and C. 
	}
\end{figure}

The hole-phonon interband scattering is strongly related to the SOC strength in semiconductors. In common semiconductors, the VBM is located at $\Gamma$ point where hh and lh bands are degenerate and above the spin-orbit split-off (SO) band by an energy separation of spin-orbit splitting $\Delta_{SO}$. Away from the $\Gamma$ point, the degeneracy of hh and lh bands is lifted by SOC with corresponding hh-lh splitting, which is proportional to $\Delta_{SO}$~\cite{harrison_elementary_1999}, along the $\Gamma-L$ and $\Gamma-X$ directions shown in Fig.~\ref{fig3}(d) for Diamond, Si, and Ge. It is straightforward to learn that the hh-lh splitting determines the number of phonons required to supply the required energy for scattering holes from hh band to lh band resulting from the energy conservation phonon scattering processes. The interband scattering rate is thus exponentially dependent on the inverse of hh-lh splitting. Subsequently, in semiconductors with a strong SOC, the interband scattering occurs only within an extremely small k range nearby the $\Gamma$ point, which means only a tiny part of holes having the opportunity to involve in the interband scattering. In striking contrast, in semiconductors with a weak SOC, most holes have the opportunity to participate in the interband scattering. Thus, Diamond, Si, InP, GaP, and AlP constituted by light (anion) atom have much stronger interband scattering owing to their small SOC than the remaining semiconductors. Such interband scattering was regarded as very ineffective in group III-V semiconductors~\cite{wiley_polar_1971}, and thus intraband scatterings are usually considered thoroughly to reproduce experimental data by adjusting the deformation potentials. Here, we demonstrate its critical role explicitly to address the abnormal low hole mobility in Si relative to Ge. 

The intraband AC scattering rate $(\tau_{hh})_{AC}^{-1}$ depends mainly on the effective DOS mass of hh band subjected to stronger phonon scatterings arising from a larger density of final states: $(\tau_{hh})_{AC}^{-1} \propto m_{d,hh}^*$. Fig.~\ref{fig3}(a) exhibits an excellent linear relationship between $(\tau_{hh})_{AC}^{-1}$ and $m_{d,hh}^*$ for all considered semiconductors except Diamond. Super hard Diamond owns a huge average elastic constant $\bar{C} \approx 86 (\times 10^{10} N/m^2)$, making AC phonons possess an exceptionally large group velocity since it is proportional to $\sqrt{\bar{C}}$. The group velocity leads large-q AC phonons to have too high frequency to be unoccupied at room temperature. This giant group velocity prevents AC phonons from the interaction with holes in Diamond, making its $(\tau_{hh})_{AC}^{-1}$ below the overall trend of other materials, as shown in Fig.~\ref{fig3}(a). It is also particularly interesting to understand why the ultrawide band-gap Diamond has an even smaller effective mass than Si. Suppose we neglect the warping of the valence bands for a moment. In this sense, we can evaluate $m_{d,hh}^*$ by inspecting $m_0/m_{hh}^{*[100]}=-1+E_P^{'}/E_0^{'} (E_0^{'}>1)$, where $E_0^{'}$ and $E_P^{'}=2Q^2/m_0$ are energy separation and energy equivalent to momentum matrix element Q between $p$-like conduction band ($\Gamma_{15c}$ for ZB) and p-like valence band ($\Gamma_{15v}$ for ZB) at $\Gamma$ point, respectively. $E_0^{'}$ scales inversely with the square of the bond length d, giving rise to $E_0^{'}=3.2$, 3.4, and 7.6 eV in Ge, Si, and Diamond. Whereas $E_0^{'}$ is usually found to be material-independent~\cite{yu_fundamentals_2010} resulting from the complementation from occupied d electrons as pointed out by Lawaetz~\cite{lawaetz_valence-band_1971}, who found that in the absence of d electrons $E_P^{'}$ scales inversely with the square of bond length. A 35\% shorter bond length in Diamond makes its $E_P^{'}$ two times bigger than Si~\cite{lawaetz_valence-band_1971}. In combination with a ratio of 0.45 in $E_P^{'}$ , it results Diamond and Si possessing similar $m_{hh}^{*[100]}$ despite their remarkable difference in band-gap. In the remaining materials, the variation in $E_P^{'}$ is not so large~\cite{lawaetz_valence-band_1971} that $m_{hh}^{*[100]}$ is highly correlated with their $E_0^{'}$. In polar semiconductors, $E_0^{'}$ depends mainly on the energy separation of cation $p$ orbital and anion $p$ orbital and is thus determined by their anion $p$ orbitals due to their cation $p$ orbitals (Al 3$p$, Ga 4$p$, and In 5$p$) are almost the same. The energy position of anion $p$ orbitals is in descending order from Sb, to As, to P, to N. Specifically, N $2p$ orbital is remarkable below remaining anion orbitals in energy (Fig.S4), giving rise to a much larger $E_0^{'}$ in III-nitrides. Thus, we expect III-nitrides to be very heavy in hh band’s effective mass, which was recently found to be the leading factor causing their exceptionally poor hole mobilities~\cite{ponce_route_2019}. From III-Sb through III-As to III-P, we deduce that the hh effective mass gets heavier and hence $(\tau_{hh})_{AC}^{-1}$ gets larger in $(\tau_{hh})_{AC}^{-1}$ with the only exceptions of AlSb and AlAs, as shown in Fig.~\ref{fig1}(b). Such exceptions are due to their significant warping of hh band due to much higher Al $3s$ orbital than remaining cation $s$ orbitals.

For a semiconductor, the optical phonon frequency governs mainly the NPO scattering since $(\tau_{hh})_{NPO}^{-1} \approx (\hbar\omega_0)^{1/2}n_0(\hbar\omega_0)$, where $n_0(\hbar\omega_0) i$s the Bose-Einstein occupation of the optical phonon with energy $\hbar\omega_0$. Fig.~\ref{fig3}(b) shows $(\tau_{hh})_{NPO}^{-1}$ against $\hbar\omega_0$ for all considered semiconductors, exhibiting a nice correlation between $(\tau_{hh})_{NPO}^{-1}$ and $\hbar\omega_0$ as indicated by a curve of $(\tau_{hh})_{NPO}^{-1} \approx (\hbar\omega_0)^{1/2}n_0(\hbar\omega_0)$. We find that Diamond has a vanishing NPO scattering rate because it has an extremely large $\hbar\omega_0=165$ meV~\cite{adachi_properties_2005}, making the optical phonons unoccupied at room temperature. Si also benefits from its sizeable $\hbar\omega_0=64$ meV to have a small intraband NPO scattering rate $(\tau_{hh})_{NPO}^{-1}$. Harrison~\cite{harrison_elementary_1999} proposed $\omega_0^2 = (127\hbar^2)/(mM_{r}d^4)$ for material MX, where $1/M_r = 1/M_M + 1/M_X$ ($M_M$ and $M_X$ are masses of the two constituent atoms M and X). Materials are expected to have simultaneously an extremely small value in reduced mass and bond length if one of its constituent atoms is of utmost light (small). It prefers elements (Be, B, N, C, O, and F) in the second row of the periodic table. Therefore, Diamond, III-Nitride and Boron-V materials have large $\hbar\omega_0$, preventing their optical phonons from participating the NPO scattering (as well as PO scattering in polar materials) in those materials.

To understand why polar material GaSb has a higher hole mobility than Si, it is crucial to reveal the atomic mechanism underlying the PO scattering in polar materials. To do so, we show in Fig.~\ref{fig3}(c) the PO scattering rate against the effective dielectric constant $\varepsilon_{\textrm{eff}}^{-1} = \varepsilon_{\infty}^{-1}-\varepsilon_s^{-1}$ ($\varepsilon_s$ is the static dielectric constant and $\varepsilon_\infty$ the high-frequency optical dielectric constant) for group III-V semiconductors. It exhibits that the PO scattering rate is highly correlated with $\varepsilon_{\textrm{eff}}$ due to it screens the macroscopic polarization field induced by the LO phonons according to Howarth~\cite{howarth_theory_1953} and Wiley~\cite{wiley_polar_1971} (Fig.S6 and Fig.S7). Because InP and GaSb have the smallest and largest $\varepsilon_{\textrm{eff}}$, their polar scattering is the strongest and weakest, respectively, among these group III-V semiconductors. Regarding the static dielectric constant of the solids is the sum of electronic and lattice contributions, $\varepsilon_s = \varepsilon_e + \varepsilon_l$ ($\varepsilon_e$ is also the optical dielectric constant $\varepsilon_\infty$), we can rewrite $\varepsilon_{\textrm{eff}}=\varepsilon_{\infty} + \varepsilon_{\infty}^{2}/\varepsilon_l$. A longer bond length enhances the electronic response to external electric field increases and screening effect on the external field. Thus, $\varepsilon_\infty$ is generally proportional to the bond length (Fig.S5). Furthermore, $\varepsilon_l$ scales with $\alpha_c^{-3}$, where $\alpha_c$ is the covalency defined by Harrison~\cite{harrison_elementary_1999}. GaSb has the smallest $s$-$s$ and $p$-$p$ energy separations between cation and anion (Fig.S4) over considered group III-V semiconductors, which render it the strongest covalency and thus the smallest $\varepsilon_l$. In combination with a sizable bond length and large $\varepsilon_\infty$ in GaSb, it causes GaSb to have the largest $\varepsilon_{\textrm{eff}}$ and the weakest polar scattering. Whereas InP holds the smallest $p$-$p$ energy separation and one of the smallest $s$-$s$ energy separations among group III-V semiconductors, rendering it to have the weakest covalency and largest $\varepsilon_l$, making InP the smallest $\varepsilon_{\textrm{eff}}$ and thus strongest polar scattering.

\textbf{Design principles towards high hole mobility}. We have learned that Si has abnormally low hole mobility because Si possesses a weak SOC resulting from the light atom and strong hole-phonon interband scattering. Compared with Ge, the weak SOC leads to small splitting in the valence bands away from $\Gamma$; therefore, it cannot suppress the interband scattering. Moreover, the short Si-Si bond enhances the coupling between $p$ orbitals to yield a larger $p$-$p$ band-gap and, thus, a heavier hole mass than Ge. On the other hand, compared with Diamond, Si is not light (small) enough to remarkably raise the optical phonons' frequency to prevent optical phonons from participating in the NPO scattering. Because the optical phonon frequency scales inversely with the square of bond length, we suggest choosing at least one of the constituent atoms from the second row in the Periodic Table (e.g., Be, B, N, C, and O) for semiconducting materials. It will own an extremely short bond length to raise the frequency of optical phonons to eliminate the NPO scattering and PO scattering (in polar materials). High-frequency optical phonons are insufficient to possess high hole mobility because the heavy hole mass and interband scattering will also deteriorate the hole mobility through the remaining AC scattering. For instance, the hole mobility of SiC ($\mu_h=98.5$\mobun) is not the average of Diamond and Si, but is a factor of 5 smaller than Si~\cite{meng_phonon-limited_2019}. This surprising behavior arises from a much heavier hole mass than both Diamond and Si regarding $\mu_{AC} \approx (m_h^*)^{5/2}$. C 2$p$ orbital is about 2 eV below the Si $3p$ orbital, which renders SiC has a larger $p$-$p$ band-gap but smaller $p$-$p$ momentum matrix element than both Diamond and Si and yields a much heavier hole mass in SiC. Therefore, we should design materials to have VBM derived from a heavy anion with strong SOC to split the valence bands and a small p-p energy separation between anion and cation to have a light hole mass. From those design principles, we can exclude nitrides~\cite{ponce_route_2019} and oxides~\cite{madelung_physics_1982}(e.g., the reported hole mobility of ZnO is {0.5-3.5 \mobun}~\cite{norton2004zno}) for high hole mobility considering much larger electronegativity of N and O than other elements, which will yield a sizable $p$-$p$ band-gap and hence heavy hole mass in compounds.  In a high-throughput computational screening, there are only very few oxides out of 3,052 oxides are predicted to have small hole mass~\cite{hautier2013identification}, which are tentatively ascribed to the hybridization between cation $s$ and O $2p$ or between O $2p$ and another anionic $p$. The low hole mobility, which is typically lower than 5 \mobun, is the main challenge impeding implementation of p-type transparent conducting oxides~\cite{yu_metal_2016}. According to our understanding, boron-containing III-V materials~\cite{liu_simultaneously_2018} are indeed expected to have high hole mobility and, more specifically, BAs should have higher hole mobility than BP due to its strong SOC. In addition to these second row elements, we should choose semiconducting materials made by cation and anion with small difference in atom size and $p$ orbital energy. The former gives rise to a longer bond length, and the latter gives rise to a light hole mass and a large effective dielectric constant, which will benefit the reduction of PO scattering.

\begin{acknowledgments}
This work was supported by the National Science Foundation for Distinguished Young Scholars under Grant No. 11925407 and the Key Research Program of Frontier Sciences, CAS under Grant No. ZDBS-LY-JSC019. S.H.W. was supported by the NSFC under Grant No. 11634003, 11991060, and U1930402 and the National Key Research and Development Program of China under Grant No. 2016YFB0700700. 
\end{acknowledgments}


\end{document}